\documentclass[a4paper,11pt]{amsart}

\usepackage[foot]{amsaddr}
\usepackage[a4paper, left=2.5cm, right=2.5cm, top=3.0cm, bottom=3.0cm]{geometry}
\usepackage[T1]{fontenc} 
\usepackage{mathrsfs}
\usepackage{graphicx}
\usepackage{subcaption}
\usepackage{tikz}
\usepackage{tabularx}
\usepackage{array}
\usepackage{graphicx,tikz} 

\title{Spectral and Phase Structure of a Unitary Matrix Model with Fisher-Hartwig Singularities}

\author{Anuj Malik}

\author{Anees Ahmed}
\address{Department of Physics, Malaviya National Institute of Technology Jaipur, Rajasthan, India 302017}
\email{anees.phy@mnit.ac.in}

\begin{document} 
	\maketitle
	

\begin{abstract}We investigate a unitary matrix model with a complex potential with Fisher-Hartwig singularities. We show that the model exhibits finite-$N$ phase transitions. The order of the phase transition is coupling-dependent. At large-$N$, these transitions are replaced by third order Gross-Witten-Wadia transitions between multiple ungapped phases and a single gapped phase, with transitions between ungapped phases forbidden. At both finite and large $N$, the phases are characterized by the locations of the Fisher-Hartwig singularities in the complex plane.
\end{abstract}

\section{Introduction}\label{sec:intro}

The study of spectral and critical properties of random matrix models has become a central interdisciplinary topic, as these models naturally arise in quantum field theory, condensed matter physics, number theory, and statistics \cite{Mehta1991, log-gases_2010, Baik2016, guhr_random-matrix_1998}. Unitary matrix models appear in a wide variety of physical systems, such as gauge theories \cite{gross_possible_1980, aharony2004, dumitru2005, hands_qcd_2010}, two-dimensional Quantum Gravity \cite{brezin_exactly_1990, douglas_strings_1990, gross_nonperturbative_1990}, the Ising model and disordered mesoscopic conductors \cite{mccoy_two-dimensional_1973, muttalib_new_1993, deift2013}, as well as topological field and string theories \cite{vleeshouwers_topological_2021}. They are also connected to the Painlevé equations and their $\tau$-functions \cite{forrester2001, szabo_two-dimensional_2012}. Simpler/limiting versions of unitary matrix models have been investigated for decades by both mathematicians and physicists; see Refs. \cite{mandal_phase_1990, russo_phases_2020, santilli_exact_2020, santilli_multiple_2022} and the references therein. We also refer to the excellent literature reviews Refs. \cite{rossi_large-n_1998} and \cite{marino_houches_2005}.

Depending on the system, the matrix model description may be non-perturbative, which has its own obvious advantages. Matrix models can serve as wonderful test labs to explore resurgent asymptotics. In fact, in certain cases the non-perturbative structure can be determined in complete detail \cite{marino2008, ahmed_transmutation_2017, Anees2018}. Additionally, matrix models have attracted attention in recent years due to their relevance to systems that exhibit the \emph{sign problem}, which obstructs the use of standard techniques such as Monte-Carlo in lattice gauge theory \cite{hands_simulating_2007}. Recent progress in resolving the sign problem has involved methods such as complex Langevin dynamics \cite{aarts2010, boguslavski2025, klauder_coherent-state_1984} and the Lefschetz-thimble approach \cite{fujii_monte_2015, cristoforetti2013, nishimura2024}.

SU($N$) matrix models are a subset of unitary matrix models, where the integral which defines the partition function is restricted to SU($N$) matrices. In standard notation, the partition function is
\begin{equation} \label{def:partitionFunction}
    Z = \int_{\text{SU}(N)} dU \, e^{- \mathcal{S}(U)/g}
\end{equation}
where $\mathcal{S}$ is the action and $g$ is the coupling. We will also use the variable $\tau=1/g$ when convenient. Expectation values in the model are defined by 
\begin{equation} \label{def:expectationValue}
    \langle h(U) \rangle = \frac{1}{Z} \int_{\text{SU}(N)} dU \, h(U) \, e^{-\mathcal{S}(U)/g}.
\end{equation}
In this work we focus on the action
\begin{equation} \label{def:action}
	\mathcal{S}(U) = - \text{Tr}\, \left(\ln(1+\alpha U)+\ln\left(1+ \frac{1}{\gamma} U^\dagger\right)\right), \qquad \alpha, \gamma > 0.
\end{equation}
As this SU($N$) model has a complex action, we expect it to violate the relation $\langle U \rangle = \langle U^\dagger \rangle$, which always holds for U($N$) models with real actions. A more interesting feature of the action is the presence of Fisher-Hartwig singularities,
and their existence implies a rich phase structure.

A generalized version of this model describes the partition function of a weakly coupled gauge theory \cite{aharony2004}, and its low temperature limit (which corresponds to $\gamma \to \infty$) has been studied in Ref.~\cite{hands_qcd_2010}. 
A U($N$) matrix model with an action similar to \eqref{def:action} famously describes the Ising spin correlation function \cite{wu1966}. In the triple-scaling limit $g \to 0, \alpha \to 0, \gamma \to \infty$ while keeping $g / \alpha$ and $ g \gamma$ equal and constant, the action \eqref{def:action}` reduces to the well-known Gross-Witten-Wadia action \cite{gross_possible_1980}. 


\section{Behaviour at finite $N$} \label{sec:finite}
The partition function \eqref{def:partitionFunction} is defined as an integral over SU($N$) matrices; that is, the integral is restricted to unitary matrices $U$ satisfying $\det U = 1 \iff \text{Tr}\, \ln U = 0$. The SU($N$) integral can be to an unconstrained U($N$) integral by means of a Fourier decomposition.
\begin{equation} \label{eq:SUNtoUN}
	\int_{\text{SU}(N)} dU \, e^{- \mathcal{S}(U)/g} = \frac{1}{2\pi} \sum_{k=-\infty}^\infty  \int_{\text{U}(N)} dU \,  e^{- \mathcal{S}(U)/g + i \, k \, \text{Tr}\, \ln U}.
\end{equation}
The SU($N$) problem is now a U($N$) problem, but at the cost of an additional infinite sum at the end. 

At finite-$N$, the U($N$) integrals may be computed by using the standard Toeplitz determinant representation
\begin{equation}
	\int_{\text{U}(N)} \, \det f(U) = \det (f_{i-j})_{i,j = 1,\dots, N}
\end{equation}
where $f_n$ are the Fourier coefficients of the symbol $f(z)$
\begin{equation}
	f_n = \oint_{|z|=1} \dfrac{dz}{2\pi i} \dfrac{f(z)}{z^{n+1}}.
\end{equation}
Combining this with \eqref{eq:SUNtoUN}, gives the following representation for the SU($N$) integral:
\begin{equation}
	\int_{\text{SU}(N)} dU \, e^{- \mathcal{S}(U)/g} = \frac{1}{2\pi} \sum_{k=-\infty}^\infty \det (f_{i-j+k})_{i,j = 1,\dots, N}
\end{equation}
With the partition function known, the free energy can be calculated as
\begin{equation} \label{def:freeEnergy}
	F = -\dfrac{1}{N^2} \ln Z
\end{equation}

The symbol associated with the action \eqref{def:action} is
\begin{equation}
	f(z) = (1+\alpha z)^\tau \left(1 + \dfrac{1}{\gamma z} \right)^\tau, \qquad \tau = 1/g.
\end{equation}
When $\tau$ is not an integer, the points $z = -1/\alpha$ and $z=-1/\gamma$ become branch points. If these branch points lie on the unit circle, then these are called the Fisher-Hartwig singularities. Thus there are four parameter regimes, defined by the positions of the branch points relative to the unit circle. As the parameter values change and the branch points cross the unit circle, the Fourier coefficients pick up contributions from the branch cuts, causing the partition function, and consequently the free energy, to have discontinuous derivatives. Of course, if $\tau \in \mathbb{Z}$ then the partition function is a smooth function everywhere. We restrict the discussion to $\gamma > 1$ and $0 < \alpha < \gamma$, as these parameter ranges are sufficient to study the behaviour described above. 

Consider first the small $\alpha$ case ($0 < \alpha < 1 < \gamma$). The branch point $z = -1/\alpha$ lies outside the unit circle while $z = -1/\gamma$ lies inside it, so no branch cut line overlaps the unit circle. The Fourier coefficients can be evaluated explicity in terms of the Gauss hypergeometric function using standard methods:
\begin{equation}
	f^\text{small}_n = u_n (\alpha, \gamma) \begin{pmatrix} \tau \\ |n| \end{pmatrix} \,_2F_1 (|n|-\tau,-\tau; 1+|n|; \alpha/\gamma) 
\end{equation}
where 
\begin{equation}
	u_n(\alpha, \gamma) = \begin{cases}
		\alpha^n, \qquad n \geq 0 \\
		\gamma^n, \qquad n \leq 0.
	\end{cases}
\end{equation}
As the Gauss hypergeometric function is analytic, $f^\text{small}_n$ are smooth functions of $\alpha$ for all $\alpha \in [0, 1]$. See Ref.~\cite{borodin2000} for an alternative expression for the small-$\alpha$ Fourier coefficients and a representation of the partition function as a Fredholm determinant.

When $\alpha$ becomes large ($1 < \alpha < \gamma$), the branch point $z = -1/\alpha$ lies inside the unit circle, and the branch cut of $(1 + \alpha z)^\tau$ intersects the unit circle. This time the Fourier coefficients pick up an additional contribution from the branch cut. The coefficients are significantly more complicated and are given by
\begin{align}
		f^\text{large}_n =f^\text{small}_n  +  \dfrac{(-1)^n}{\pi} \dfrac{\sin (\pi \tau)}{\tau+1} \alpha^n (\alpha-1)^{\tau+1}  \left(1 - \dfrac{\alpha}{\gamma} \right)^\tau \, A(n, \tau, \alpha, \gamma)
\end{align}
where $A(n, \tau, \alpha, \gamma)$ is shorthand for the Appell function
\begin{equation}
	A(n, \tau, \alpha, \gamma) := F_1 \left(\tau+1; n+\tau+1, -\tau;\tau+2;1-\alpha, \dfrac{1-\alpha}{1-\alpha/\gamma} \right).
\end{equation}
The term containing this function is the contribution from the branch cut, and it vanishes if $\tau \in \mathbb{Z}$. While the Appell function is infinitely differentiable near $\alpha=1$, the pre-factor is not. The behaviour of the branch cut contribution near $\alpha = 1$ is 
\begin{equation}
	(-1)^n \dfrac{\sin (\pi \tau)}{\pi ( 1 + \tau)} \, \left(1 - 1/\gamma \right)^\tau \, \epsilon^{\tau + 1}, \qquad \epsilon = \alpha - 1.
\end{equation}
It is clear that there is a discontinuity in the $k$-th derivative if $\tau \neq \mathbb{Z}$, where $k = 1 + \mathrm{ceiling}(\tau)$, and this discontinuity propagates to the partition function and free energy. Therefore, in the Ehrenfest classification, the order of the phase transition  at $\alpha=1$ is $1 +\mathrm{ceiling}(\tau)$.

\begin{figure}[t]
	\centering
	\includegraphics[width=0.7\linewidth]{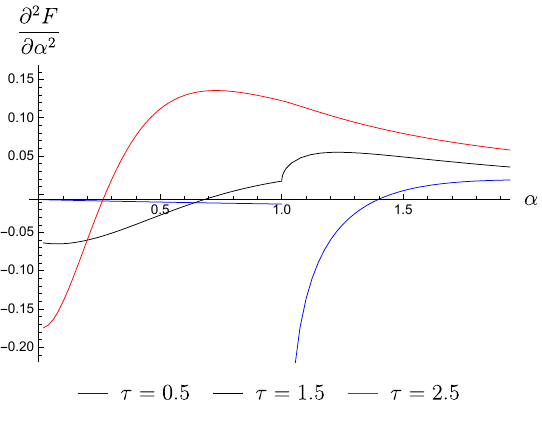}
	\caption{\small Second derivative of the free energy of the U($N$) model with respect to $\alpha$ at $N=3$.}
	\label{fig:finiteFreeEnergy}
\end{figure}

With the Fourier coefficients calculated, the finite-$N$ partition function and free energy can be studied. We leave the full expressions of these quantities for the SU($N$) model for a future work. Fig.~\ref{fig:finiteFreeEnergy} shows the second derivative of the $N=3$ free energy of the U($N$) model with respect to $\alpha$. It is clear that for $\tau=0.5$ and $1.5$, the order of transition is $2$ and $3$ respectively, consistent with the discussion above.

In the large-$N$ limit, the coupling $g$ must be scaled inversely with $N$, otherwise the model reduces to the trivial model $Z = \int dU$. Thus, $\tau \propto N$ in the large-$N$ limit, and consequently, the order of the phase transition approaches infinity. This indicates that the large-$N$ partition function becomes smooth at $\alpha = 1$ (and also $\gamma = 1$). In other words, the finite-$N$ phase transition disappears in the large-$N$ limit. In the next section we work in the exact large-$N$ limit and show that this is indeed the case.

\section{Large-$N$ limit} \label{sec:largeN}
We work in the double scaling limit $N \to \infty, g \to 0$ while keeping constant the following combination
\begin{equation}
	\sigma = \dfrac{1}{N g}.
\end{equation}
We define the potential $V$ by
\begin{equation}
	\mathcal{S}(U)/g = N \, \mathrm{Tr}\, V(U).
\end{equation}
The potential corresponding to the action \eqref{def:action} is
\begin{equation} \label{def:potential}
	V(z) = - \sigma \left(\ln(1+\alpha z)+\ln\left(1+ \frac{1}{\gamma z}\right)\right).
\end{equation}
In the exact large-$N$ limit the SU($N$) integral with potential $V(z)$ is equivalent to a U($N$) integral with the potential $V(z) + \mathscr{N} \ln z$ \cite{hands_qcd_2010}. Here $\mathscr{N}$ is a Lagrange multiplier that enforces the SU($N$) constraint $\det U = 1$, and it must be determined self-consistently by demanding $\langle \mathrm{Tr}\, \ln U \rangle = 0$.

The standard procedure now is to introduce the eigenvalues $e^{i\theta_i}$ ($i=1,\ldots,N$) of $U$, and rewrite the U($N$) integral as an integration over the angular variables $\theta_i \in [-\pi,\pi)$ as
\begin{equation}  \label{eq:partitionFunctionAsAngularIntegral}
    Z = \int \Big(\prod_{i = 1}^N d\theta_i \Big) \, e^{-\mathcal{S}_\text{eff}/g},
\end{equation}
where $\mathcal{S}_\text{eff}$ is the effective action given by
\begin{equation} \label{eq:effectiveAction}
    \mathcal{S}_{\text{eff}}/g = N \sum_{i=1}^{N} V(e^{i \theta_i}) -\frac{1}{2}\sum_{i,j=1}^{N}\ln{\sin^2{\left(\frac{\theta_i-\theta_j}{2}\right)}}  + i N \mathscr{N} \sum_{i=1}^{N} \theta_i.
\end{equation}
The second sum in the above expression is the Vandermonde contribution. 
In this form, an approximation to the integral can be found by using the saddle point method. As we are working in the exact large-$N$ limit, this approximation is actually exact. The saddle point equation, $\partial \mathcal{S}_\text{eff} / \partial \theta_i = 0$ for every $i$, works out to be 
\begin{equation}
    V'(\theta_i) + i\mathscr{N}  = \frac{1}{N}\sum_{i\neq j}\cot{\left(\frac{\theta_i-\theta_j}{2}\right)}.
\end{equation}
It is clear from the form of the equation that the saddles may be complex.

The next step is to introduce the spectral function, defined formally as
\begin{equation} \label{def:angularDensity}
    \bar{\rho}(\theta) = \frac{1}{N} \sum_i \delta(\theta - \theta_i),
\end{equation}
which allows the following replacement when taking the large-$N$ limit:
\begin{equation}
    \frac{1}{N}\sum_{i}^N f(\theta_i) \to \int d\theta \, \bar{\rho}(\theta) f(\theta).
\end{equation}
$\bar{\rho}(\theta)$ is normalized to unity by virtue of its definition \eqref{def:angularDensity}. We define $z_i = e^{i \theta_i}$ to return to the space of eigenvalues of $U$. The saddle point equation now takes the form
\begin{equation}
    z_iV'(z_i) + \mathscr{N}= \frac{1}{N}\sum_{i\neq j}\frac{z_i+z_j}{z_i-z_j}
\end{equation}
and its large-$N$ limit is 
\begin{equation} \label{EOM}
    zV'(z) + \mathscr{N} = \mathscr{P} \int_\mathcal{C} \frac{dz'}{2 \pi i z'} \, \rho(z') \, \frac{z + z'}{z-z'}, \qquad z \in \mathcal{C}.
\end{equation}
In the above relation $\mathscr{P}$ denotes the principal value of the integral and $\rho(z) := 2 \pi \bar{\rho}(\theta)$ for all $z \in \mathcal{C}$, where $\mathcal{C}$ is the contour on which the saddle points coalesce (also called the support of $\rho(z)$). The density of the saddle points on $\mathcal{C}$ is 
\begin{equation}
	D(z) = \dfrac{\rho(z)}{2 \pi i z}, \qquad z \in \mathcal{C}.
\end{equation}
The normalization of $\rho(z)$ follows from that of $\bar{\rho}(\theta)$:
\begin{equation} \label{eq:spectralFunctionNormalization}
    \int_C\dfrac{dz}{2\pi i z} \, \rho(z)= 1.
\end{equation}
We continue $\rho(z)$ to the entire complex plane by using \eqref{EOM} as its definition.

The problem now is to find the spectral function $\rho(z)$ and its support $\mathcal{C}$ given the potential \eqref{def:potential}. This is a Reimann-Hilbert problem \cite{brezin1978, marino_houches_2005}, and it can be solved in a rather straight-forward manner if we restrict to solutions in which $\mathcal{C}$ is a single continuous curve. Such a solution is called a single cut solution. For U($N$) matrix integrals with real action the saddle points are found exclusively on the unit circle and $\mathcal{C}$ is simply the unit circle or an arc of a unit circle. As the present model is an SU($N$) integral with a complex action, the saddle points may be found off the unit circle. Once $\rho(z)$ is known, $\mathcal{C}$ can be found by solving
\begin{equation}\label{Contour}
    \dfrac{dz}{d\theta}  = \frac{i z}{\rho(z)}, \qquad z \equiv z(\theta).
\end{equation}
The expression for the expectation values simplifies drastically in the large-$N$ limit:
\begin{equation}
    \frac{1}{N} \, \langle \mathrm{Tr} \, h(U) \rangle \to  \int_C\dfrac{dz}{2\pi i z} \rho(z) \, h(z).
\end{equation}
Lastly, the spectral function must satisfy the SU($N$) constraint, which we can derive as follows:
\begin{equation}\label{SU}
   \det U = 1 \implies \sum_i \theta_i = 0 \implies \int_C\frac{dz}{2\pi i z}\rho(z)\ln{z}=0.
\end{equation}
Next we discuss the ungapped phases of the model.

\subsection{Ungapped phases}  \label{sec:confinedPhase}

\begin{figure}[t]
    \centering
    \includegraphics[width=0.49\linewidth]{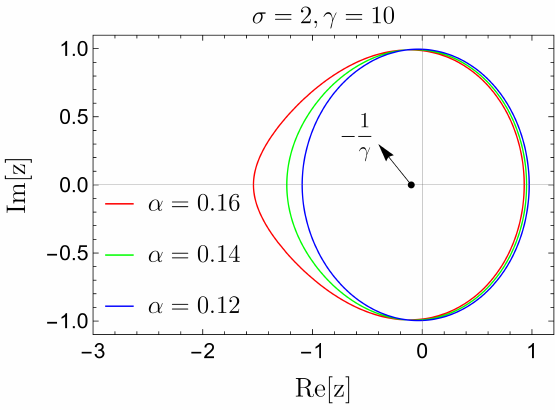}
     \includegraphics[width=0.49\linewidth]{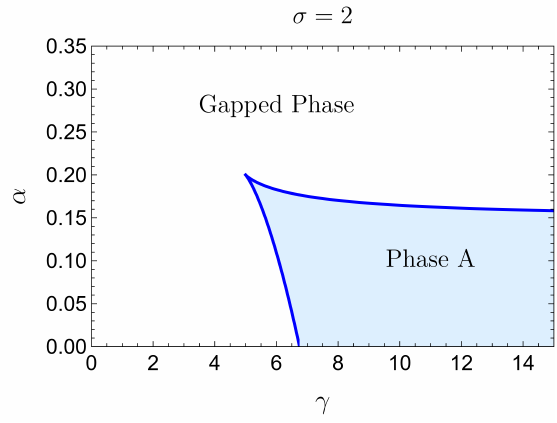} 
    \caption{\small Left: support of the spectral function in the ungapped phase A (small $\alpha$ large $\gamma$ phase) for a few values of $\alpha$. The dot at $z=-1/\gamma$ is the pole which is inside $\mathcal{C}$, while the other pole (not shown) is outside. Right: phase boundary between the ungapped phase A and the gapped phase.}
    \label{fig:phaseA}
\end{figure}

\begin{figure}[t]
    \centering
    \includegraphics[width=0.50\linewidth]{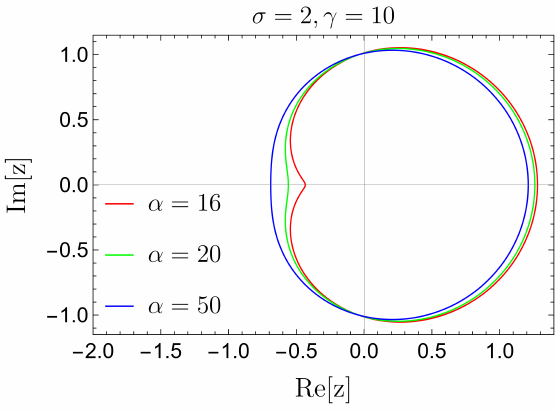}
    \includegraphics[width=0.48\linewidth]{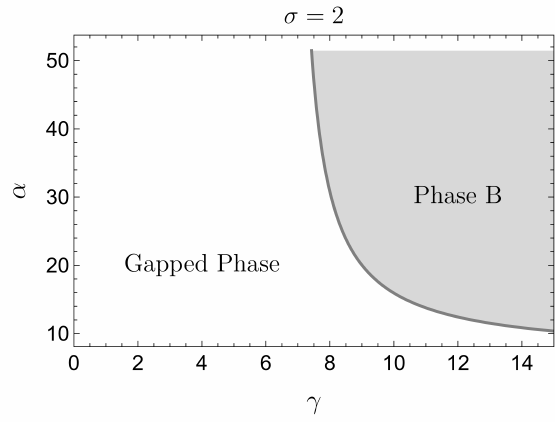} \\
    \includegraphics[width=0.49\linewidth]{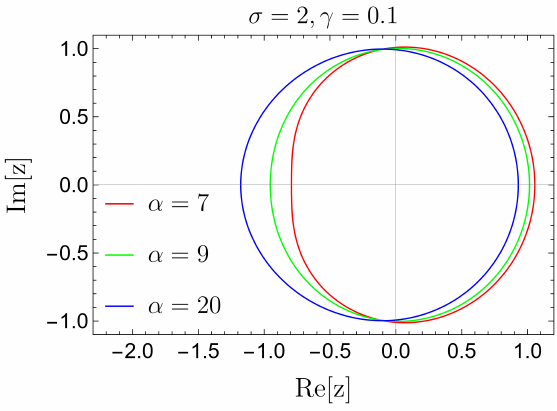}
    \includegraphics[width=0.49\linewidth]{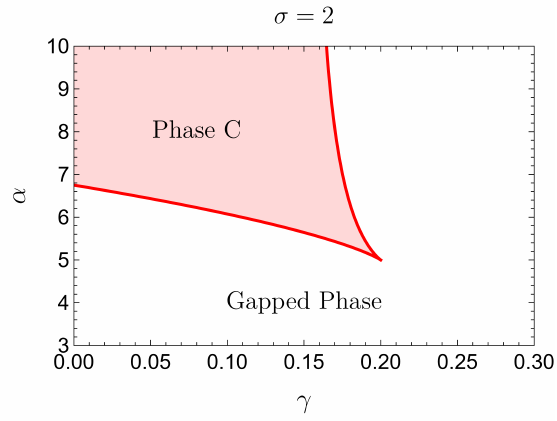} \\
    \includegraphics[width=0.49\linewidth]{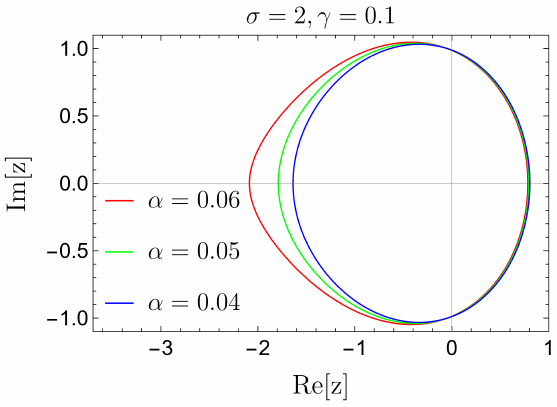}
    \includegraphics[width=0.50\linewidth]{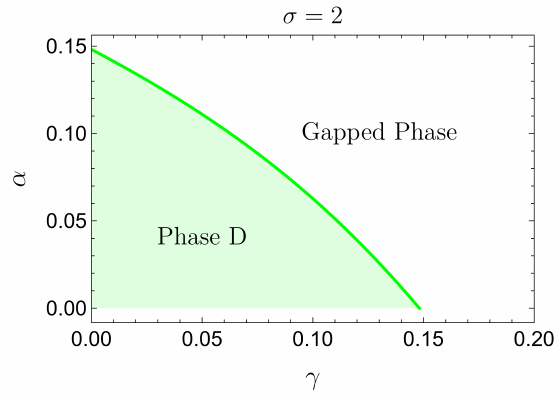}
    \caption{\small Support of the spectral function (left panels) and the phase boundary between the gapped and ungapped phases (right panels). The panels from top to bottom correspond to ungapped phases B, C and D in the same order. }
    \label{fig: PhaseB}
\end{figure}

We refer to any phase in which the spectral function has a closed support as an ungapped phase of the model. In this phase all integrals over the support, such as the saddle point equation \eqref{EOM}
\begin{equation} \label{eq:ungappedSaddleEquation}
    \sigma \left(- 1 + \dfrac{1}{1+\alpha z} + \dfrac{1}{1+ \gamma z}\right) + \mathscr{N} = \mathscr{P} \oint_\mathcal{C} \frac{dz'}{2\pi i z'} \, \rho(z') \, \frac{z + z'}{z-z'}, \qquad z \in \mathcal{C},
\end{equation}
become closed contour integrals. Thus, we can immediately utilize elementary theorems to solve the model and compute various quantities of interest.

We assume the following ansatz based on the branch points of the potential ($z = -\frac{1}{\alpha}, -\frac{1}{\gamma}$):
\begin{equation} \label{eq:spectralAnsatz}
	\rho(z) = \sum_{n=0}^{\infty} \frac{A_n}{z^n} + \sum_{n=0}^{\infty} \frac{B_n}{(z + 1/\alpha)^n} + \sum_{n=0}^{\infty} \frac{C_n}{(z + 1/\gamma)^n}.
\end{equation}
It is necessary to include a pole at $z=0$ in the ansatz to ensure that the normalization condition \eqref{eq:spectralFunctionNormalization} is satisfied, and this pole must always lie inside $\mathcal{C}$. The coefficients in the ansatz are determined by substituting it into \eqref{eq:ungappedSaddleEquation} and comparing both sides. In addition, the normalization condition may be used to determine any one coefficient.

There are four different solutions for $\rho(z)$ distinguished by whether the poles $z = -\frac{1}{\alpha}$ and $z=-\frac{1}{\gamma}$ are inside or outside $\mathcal{C}$. Each of these solutions corresponds to a distinct ungapped phase. Just as in the finite-$N$ case, the phases of the model are characterized by the branch points of the potential. There is, however, a critical difference between the finite-$N$ and large-$N$ transitions, which we will discuss at the end. We use the following nomenclature henceforth:
\begin{itemize} \itemsep0pt
    \item Phase A: Only the pole $z = -1/\gamma$ lies inside $\mathcal{C}$
    \item Phase B: Both the poles lie inside $\mathcal{C}$
    \item Phase C: Only the pole $z = -1/\alpha$ lies inside $\mathcal{C}$
    \item Phase D: No poles lie inside $\mathcal{C}$
\end{itemize}

For brevity, we discuss phase A in detail and summarize the rest at the end of the section. This phase can be thought of as the small $\alpha$ large $\gamma$ phase. We substitute the ansatz \eqref{eq:spectralAnsatz} in \eqref{eq:ungappedSaddleEquation} and compare sides to obtain the spectral function \begin{equation} \label{eq:densityA}
    \rho(z) = 1  +\sigma+\frac{\sigma}{1+\gamma z}-\frac{\sigma}{1+\alpha z}.
\end{equation}
The comparison also shows that the Lagrange multiplier $\mathscr{N}$ vanishes. With the spectral function known, the SU($N$) constraint \eqref{SU} simplifies to
\begin{equation}\label{eq:phaseA_SU_condition}
    (1 + \sigma) \ln{r_0} + \sigma \ln{\left( \frac{1 - r_0 \alpha }{ r_0 \gamma - 1 } \right)}  + \sigma \ln \gamma= 0
\end{equation}
where $r_0 = |z(\pm \pi)|$ is the distance to the intersection of $\mathcal{C}$ and the negative real axis. Since the pole at $z = -\frac{1}{\alpha}$ is outside $\mathcal{C}$, the number $r_0$ is constrained to $\frac{1}{\gamma} < r_0 < \frac{1}{\alpha}$. This number serves as an initial value for the differential equation \eqref{Contour}, which can be solved to obtain the equation for the support $\mathcal{C}$:
\begin{equation} \label{eq:contourA}
    \alpha|z|^{1+1/\sigma} = \left|z + \frac{1}{\alpha} \right|^{-1} \, \left|z +\frac{1}{\gamma} \right|.
\end{equation}
As expected, $\mathcal{C}$ is not a unit circle in this phase. See the left panel in Fig. \ref{fig:phaseA} for plots of $\mathcal{C}$ in phase A. 

\begin{figure}[t]
	\centering
	\includegraphics[width=.47\textwidth]{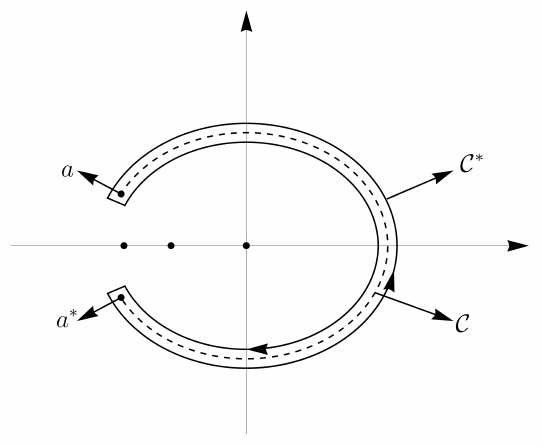}
	\qquad
	\includegraphics[width=.47\textwidth]{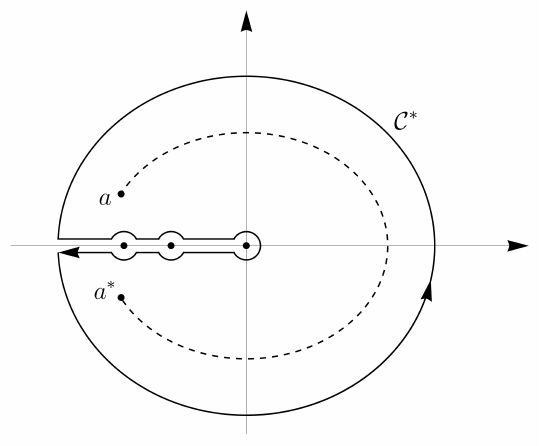}
	\caption{\small Schematic of the support $\mathcal{C}$ of the spectral function in the gapped phase (dashed, both panels). The dots on the negative real axis are the branch points of the potential ($z = -1/\alpha, -1/\gamma$). Solid closed curve in the left panel is the contour $\mathcal{C}^*$ used to solve integrals in the gapped phase. It can be deformed continuously to that shown in the right panel. See Fig. \ref{fig:gappedContour} for the actual contours in the gapped phase.}
	\label{fig: deformed contour}
\end{figure}

If for some value of $\alpha$ the spectral function acquires a zero on the negative real axis, say $z = -r_0$, then we have a phase transition to another phase. We call this new phase the \emph{gapped} phase as $\mathcal{C}$ has acquired a gap at $z=-r_0$. Solving $\rho(-r_0) = 0$ gives the phase boundary
\begin{equation} \label{eq:boundaryA}
    \alpha = \frac{\gamma  r_0-\sigma -1}{r_0 (\gamma  r_0 (\sigma +1)- 2 \sigma - 1)}
\end{equation}
where $r_0$, given by \eqref{eq:phaseA_SU_condition}, is itself a function of $\alpha$ and $\gamma$. We solve this equation numerically and plot the phase diagram close to the transition between the ungapped phase A and the gapped phase in the right panel of Fig.~\ref{fig:phaseA}.

Next, we calculate the moments in this phase.
\begin{equation}
    W_n := \frac{1}{N} \langle \mathrm{tr}\, U^n \rangle =  \oint_C\dfrac{dz}{2\pi i z} \rho(z) z^n = \begin{cases}  
    (-1)^{n+1}\frac{\sigma}{\gamma^n} \hspace{1.5cm} n\geq 1\\
     (-1)^{n+1}\sigma\alpha^{|n|} \hspace{1.2cm} n\leq -1
    \end{cases}
\end{equation}
If our model results from a lattice gauge theory, then these moments are related to the winding Wilson loops over a single plaquette \cite{okuyama2017}. We observe that 
\begin{equation}
    W_n \neq W_{-n}
\end{equation}
which indicates the presence of the sign problem. This behavior is completely expected, as our model's potential is complex, in addition to having an SU($N$) geometry. 

Lastly, we look at the free energy $F$, defined in \eqref{def:freeEnergy}. Since $F$ in the large-$N$ limit is just the rescaled effective action, $F = \frac{\sigma}{N} \mathcal{S}_\text{eff}$, the continuum version of \eqref{eq:effectiveAction} can be used to directly calculate the free energy. Another method is to first find the derivatives of the free energy with respect to each parameter of the model using
\begin{equation}
    \frac{\partial F}{\partial t} = \left \langle \frac{\partial V(U)}{\partial t} \right \rangle =  \oint_C   \frac{dz }{2\pi i z} \rho(z) \frac{\partial V(z)}{\partial t}, \qquad t = \sigma,\alpha, \gamma,
\end{equation}
and then integrate back. The derivatives are
\begin{align}
    \frac{\partial F}{\partial \alpha} =  \frac{\sigma^2}{\alpha - \gamma}, \qquad \frac{\partial F}{\partial \gamma} = -\frac{\sigma^2}{\gamma} + \frac{\sigma^2}{ \gamma-\alpha}, \qquad \frac{\partial F}{\partial \sigma} =  2 \sigma \ln{\left(1-\frac{\alpha}{\gamma}\right)}.
\end{align}
Using either method, we find the free energy in phase A to be
\begin{equation}
    F =   \sigma^2 \ln \left( 1-\frac{\alpha}{\gamma} \right).
\end{equation}
In the tables below we summarize the important quantities in each phase. See Fig.~\ref{fig:phaseA} and Fig.~\ref{fig: PhaseB} for plots of the support of the spectral function for each ungapped phase and the phase diagram near the ungapped-gapped transition.

\begin{center}
\renewcommand{\arraystretch}{2.0}
\begin{tabular}{|l|p{5cm}|p{5cm}|}
    \hline
    & \textbf{Phase A} & \textbf{Phase B} \\
    & (small $\alpha$ large $\gamma$ phase) & (large $\alpha$ large $\gamma$ phase) \\
    \hline
    Spectral function $\rho(z)$ & $1  +\sigma+\dfrac{\sigma}{1+\gamma z}-\dfrac{\sigma}{1+\alpha z}$  & $1+\dfrac{\sigma}{1+\gamma z}+\dfrac{\sigma}{1+\alpha z}$\\
    \hline
    Support $\mathcal{C}$ & $ \alpha |z|^{1+1/\sigma} = \Big|z + \dfrac{1}{\alpha}\Big|^{-1} \, \Big|z +\dfrac{1}{\gamma}\Big| $ & $ |z|^{2 + 1/\sigma} = \Big|z + \dfrac{1}{\alpha}\Big| \, \Big|z+\dfrac{1}{\gamma}\Big| $ \\
    \hline 
    Lagrange multiplier $\mathscr{N}$ & 0 & $\sigma$ \\
    \hline
    Wilson loop $W_{n \geq 1}$ & $ (-1)^{n+1} \sigma/ \gamma^n$ & 
    $(-1)^{n+1}\sigma\left(\dfrac{1}{\alpha^n}+\dfrac{1}{\gamma^n}\right)$ \\
    \hline
    Wilson loop $W_{n \leq -1}$ & $ (-1)^{n+1}\sigma/\alpha^n$ & $0$  \\
    \hline 
    Free energy $F$ & $\sigma^2 \ln{\left( 1 - \dfrac{\alpha}{\gamma} \right)}$ & $-\sigma \ln \alpha$ \\
    \hline
\end{tabular}
\end{center}

\begin{center}
\renewcommand{\arraystretch}{2.0}
\begin{tabular}{|l|p{5cm}|p{5cm}|}
    \hline
    & \textbf{Phase C} & \textbf{Phase D} \\
    & (large $\alpha$ small $\gamma$ phase) & (small $\alpha$ small $\gamma$ phase) \\
    \hline
    Spectral function $\rho(z)$ & $1+\sigma +\dfrac{\sigma}{1+\alpha z}-\dfrac{\sigma }{1+\gamma z}$  & $1 + 2\sigma - \dfrac{\sigma}{1 + \gamma z} - \dfrac{\sigma}{1 + \alpha z}$\\
    \hline
    Support $\mathcal{C}$ & $ \gamma |z|^{1+1/\sigma} = \Big|z + \dfrac{1}{\alpha}\Big| \, \Big|z + \dfrac{1}{\gamma}\Big|^{-1} $ & $ \alpha \gamma |z|^{1/\sigma} = \Big|z + \dfrac{1}{\alpha}\Big|^{-1} \, \Big|z + \dfrac{1}{\gamma}\Big|^{-1}$\\
    \hline 
    Lagrange multiplier $\mathscr{N}$ & 0 & $-\sigma$ \\
    \hline
    Wilson loop $W_{n \geq 1}$ & $ (-1)^{n+1} \sigma/\alpha^n $ & $0$\\
    \hline
     Wilson loop $W_{n \leq -1}$ & $(-1)^{n+1}\sigma/\gamma^n $ & $(-1)^{n+1}\sigma\left(\dfrac{1}{\alpha^n}+\dfrac{1}{\gamma^n}\right)$  \\
    \hline 
    Free energy $F$ & $\sigma ^2 \ln{\left( 1 - \dfrac{\gamma}{\alpha} \right)}  +\sigma  \ln{\dfrac{\gamma}{\alpha}}$ & $\sigma \ln{\gamma}$\\
    \hline
\end{tabular}
\end{center}

\subsection{Gapped Phase}\label{sec:GappedPhase}
\begin{figure}[t]
	\centering
	\includegraphics[width=0.6\linewidth]{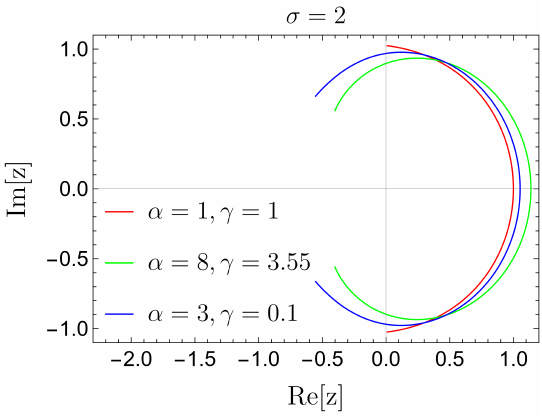}
	\caption{\small Support of the spectral function $\rho(z)$ in the gapped phase.}
	\label{fig:gappedContour}
\end{figure}
In this phase the saddle points lie on an open contour. The \emph{gap} develops about the negative real axis, similar to that in the Gross-Witten-Wadia model. The gapped phase can be studied by using the resolvent, which is defined as
\begin{equation}
    R(z) = \left\langle \frac{z+U}{z-U} \right\rangle .
\end{equation}
This definition implies the limiting values
\begin{equation}\label{endpoint}
    R(z)=\begin{cases}
        -1 \hspace{1cm}  z\rightarrow0,\\
        +1 \hspace{1cm}   z\rightarrow\infty,\\
    \end{cases}
\end{equation}
and in the large-$N$ limit the resolvent can be expressed as an integral over the support $\mathcal{C}$
\begin{equation}\label{Resolvent}
    R(z) = \int_\mathcal{C} \frac{dz'}{2\pi i z'}\,\rho(z')\,\frac{z+z'}{z-z'}.
\end{equation}
It is clear from the above expression that $R(z)$ is analytic everywhere except for $z \in \mathcal{C}$, and that it has branch points at the ends of $\mathcal{C}$. We denote the ends of $\mathcal{C}$ by $a$ and $a^*$.

\begin{figure}[t]
	\centering
	\includegraphics[width=.49\textwidth]{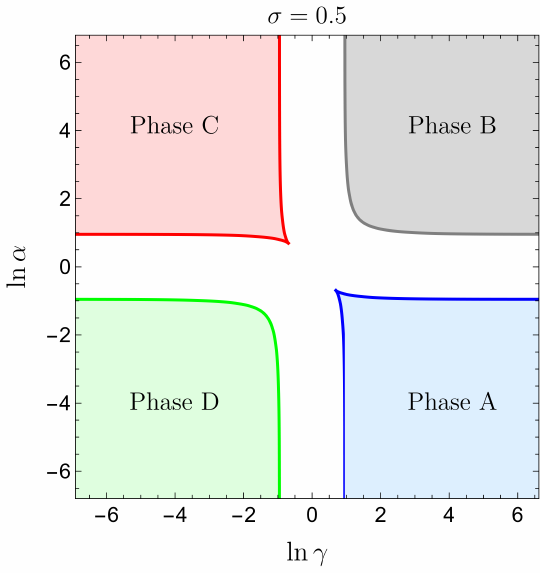}
	\,
	\includegraphics[width=.49\textwidth]{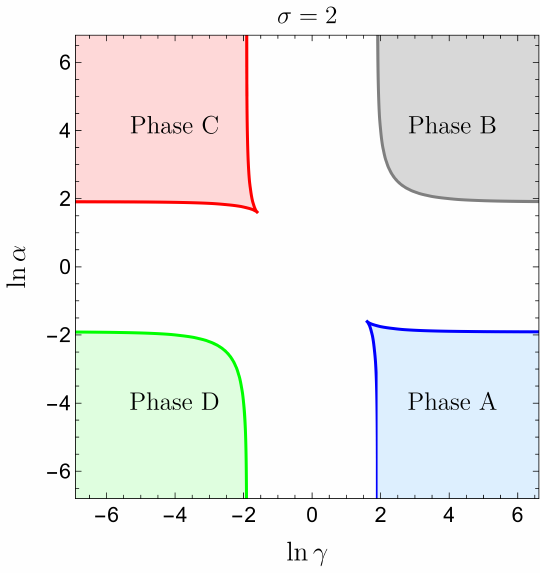}
	\caption{\small Large-$N$ phase diagram of the model for $\sigma=0.5$ and $\sigma=2$. The four ungapped phases are represented by various colors and the gapped phase by white. As $\sigma$ is increased the gapped phase grows outwards from the center $\alpha=1, \gamma=1$.}
	\label{fig: Phase diagram}
\end{figure}

By using the Sokhotski-Plemelj formula \cite{muskhelishvili2013}, we find the two main expressions which make the resolvent a useful quantity. The first relates the potential to the resolvent:
\begin{equation}\label{CutEOM}
    zV'(z) = \frac{1}{2} \, \lim_{\epsilon\to 0}\Big(R(z+\epsilon)+R(z-\epsilon)\Big).
\end{equation}
The second is an expression for the spectral function in terms of the resolvent:
\begin{equation}\label{eq:densityFromResolvent}
    \rho(z)= \frac{1}{2} \Delta_\mathcal{C} R(z)
\end{equation}
where $\Delta_\mathcal{C} R(z)$ is the discontinuity in the resolvent across $\mathcal{C}$.

Motivated by \eqref{CutEOM} and \eqref{eq:densityFromResolvent}, and the fact that $\mathcal{C}$ can serve as a branch cut for $R(z)$, we write the following ansatz for the resolvent:
\begin{equation}
    R(z) = z V'(z)+ g(z)Q(z).
\end{equation}
The second term contains $g(z)$, an unknown function well-behaved on $\mathcal{C}$ but possibly meromorphic outside it, and $Q(z)$, which contains the discontinuity in $R(z)$. The latter is defined as $Q(z) = \sqrt{z-a} \, \sqrt{z-a^*}$ with branch cut $\mathcal{C}$. This definition ensures that $\rho(a) = \rho(a^*) = 0$. Using the ansatz in \eqref{eq:densityFromResolvent} gives 
\begin{equation}
    \rho(z) = \frac{1}{2} g(z) \, \Delta_\mathcal{C} Q(z).
\end{equation}
Thus, the problem of finding the spectral density is equivalent to the problem of finding $g(z)$, the Lagrange multiplier $\mathscr{N}$ and the endpoints $a, a^*$.

\begin{figure}
    \centering
    \includegraphics[width=.49\textwidth]{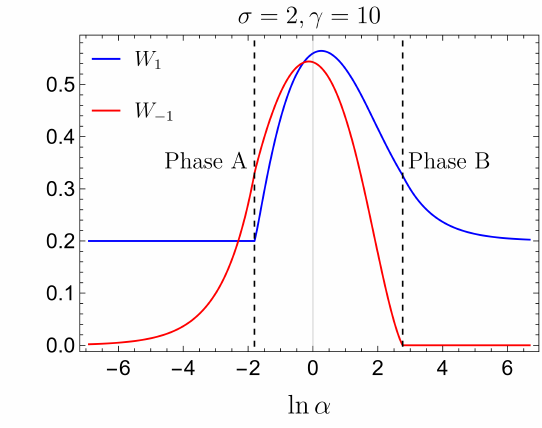}
    \,
    \includegraphics[width=.49\textwidth]{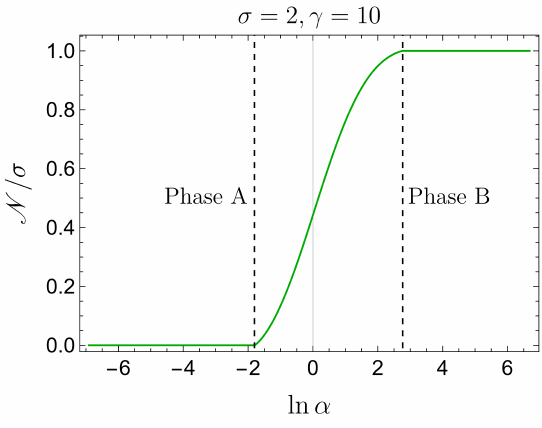} \\
    \quad \\
    \includegraphics[width=.49\textwidth]{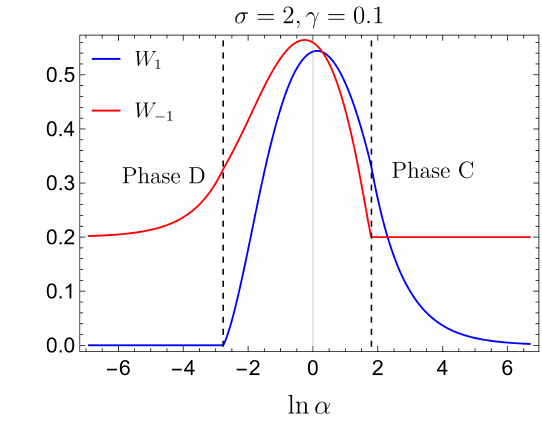}
    \,
    \includegraphics[width=.49\textwidth]{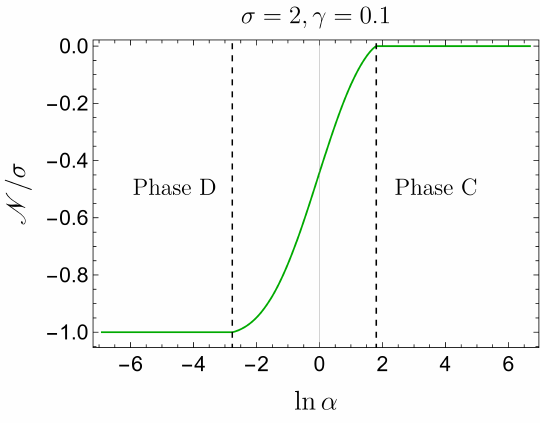}
    \caption{\small Wilson loop (left panels) and Lagrange multiplier (right panels) in the large-$N$ limit as the system transitions from ungapped phase A to B (top panels) and from phase D to C (bottom panels) through the gapped phase.}
    \label{fig:gappedWilson}
\end{figure}

To find an expression for $g(z)$, we first note that \eqref{eq:densityFromResolvent} allows writing the expectation value of a quantity as a closed contour integral:
\begin{equation} \label{eq:openToClosed}
    \frac{1}{N} \langle \mathrm{Tr} \,h(U) \rangle = \int_\mathcal{C} \dfrac{dz}{2\pi i z} \, \rho(z) \, h(z)   = \oint_{\mathcal{C}^*} \dfrac{dz}{4 \pi i z} \, R(z) \, h(z), 
\end{equation}
where $\mathcal{C}^*$ is a closed contour wrapped around the support $\mathcal{C}$, as shown in the left panel of Fig \ref{fig: deformed contour}. Then, a general expression for $g(z)$ is 
\begin{equation}\label{g(z)}
   g(z)= \oint_{\mathcal{C}^*}\frac{z' V(z')}{(z'-z)Q(z')}\frac{dz'}{2\pi i}.
\end{equation}
This result is similar to a theorem by Tricomi \cite{tricomi1985}. For the current model, $\mathcal{C}^*$ can be continuously deformed to that shown in the right panel, which means that the integral in the above relation takes contributions only from the residues at $z=0,-\frac{1}{\alpha},-\frac{1}{\gamma}, \infty$. Solving the integral we find
\begin{equation}
    g(z)= \frac{\sigma}{(1+\gamma z) B_\gamma} + \frac{\sigma}{(1+\alpha z) B_\alpha}
\end{equation}
where we have used the shorthands
\begin{equation*}
    B_\alpha = \left|\frac{1}{\alpha}+a\right|, \quad B_\gamma = \left|\frac{1}{\gamma}+a\right|.
\end{equation*}

For the remaining three quantities $a$, $a^*$ and $\mathscr{N}$, we need three independent equations. As $g(z)$ is now known, two equations are found immediately from the limiting behaviour of the resolvent as given in \eqref{endpoint}.

\begin{equation}\label{endpointConstraints}
    \mathscr{N}+\sigma-\sigma |a| \left(\frac{1}{B_\alpha}+\frac{1}{B_\gamma}\right) =-1, \quad \mathscr{N}-\sigma+\sigma\left(\frac{1}{\alpha \, B_\alpha}+\frac{1}{\gamma B_\gamma}\right) =1.
\end{equation}
The third and last equation is found by enforcing the SU($N$) constraint $\langle \mathrm{Tr} \,\ln U \rangle = 0$. By using \eqref{eq:openToClosed} this constraint can be rewritten in the following form 
\begin{equation}
    \oint_{\mathcal{C}^*}  \dfrac{dz}{4 \pi i z} \, R(z) \, \ln z = 0
\end{equation}
By deforming the contour $\mathcal{C}^*$ as shown in Fig. \ref{fig: deformed contour}, we find the SU($N$) constraint to be
\begin{equation}\begin{aligned}
    & \dfrac{1}{B_\alpha} \left[ \left( |a| + \dfrac{1}{\alpha} \right) \ln |a| - \left( |a| - \dfrac{1}{\alpha} \right) \ln \cos^2 \frac{\phi}{2} \right] -  \text{arcsinh} \dfrac{(\alpha - |a|^{-1}) B_\alpha}{1 + \cos \phi} + \ln \alpha\\
    & + \dfrac{1}{B_\gamma} \left[ \left( |a| + \dfrac{1}{\gamma} \right) \ln |a| - \left( |a| - \dfrac{1}{\gamma} \right) \ln \cos^2 \frac{\phi}{2} \right] -  \text{arcsinh} \dfrac{(\gamma - |a|^{-1}) B_\gamma}{1 + \cos \phi} + \ln \gamma = 0,
\end{aligned}\end{equation}
where $\phi\in (0, \pi)$ is the argument of the endpoint $a$. Simultaneously solving the equations \eqref{endpointConstraints}, 
together with the SU($N$) constraint gives the Lagrange multiplier $\mathscr{N}$ and the endpoints $a,a^*$.

Due to the transcendental nature of the SU($N$) constraint, the observables in this phase can not be expressed in closed-form in terms of the model parameters $\sigma, \alpha, \gamma$. For example, the single-winding Wilson loop expectation values
\begin{align}
   &W_1 =\dfrac{\sigma}{2}\left(\dfrac{1}{\alpha }+\dfrac{1}{\gamma }\right)-\dfrac{\sigma}{2 |a \alpha + 1|} \left(\dfrac{1}{\alpha} + \text{Re}\, a \right) - \dfrac{\sigma}{2 |a \gamma  + 1|} \left(\dfrac{1}{\gamma} + \text{Re}\, a \right) \\
   &W_{-1} = \dfrac{\sigma}{2}(\alpha +\gamma )-\dfrac{\sigma \alpha }{2 |\alpha + 1/a|} \left( \alpha  + \dfrac{\text{Re}\, a}{|a|^2} \right) - \dfrac{\sigma \gamma }{2 |\gamma + 1/a|} \left( \gamma  + \dfrac{\text{Re}\, a}{|a|^2} \right)
\end{align}
depend on the endpoints $a,a^*$ which themselves depend transcendentally on the model parameters. The situation with the free energy is identical. Here we provide the derivative of free energy with respect to $\alpha$ as an example.

\begin{equation}
      \frac{\partial F}{\partial \alpha}= -\frac{\sigma^2}{2\alpha} \left(\frac{1}{\alpha B_\alpha} + \frac{1}{\gamma B_\gamma}\right) + \frac{\sigma ^2}{2(\alpha -\gamma) }\left(1-\frac{B_\gamma}{B_\alpha} \right)   +\frac{\sigma^2}{2\alpha^2 A^2_\alpha}\left(\frac{1}{\alpha} + \text{Re} \, a \right).
\end{equation}
The support, found by numerically solving \eqref{Contour}, is shown in Fig. \ref{fig:gappedContour}.  All plots for the gapped phase were made using Mathematica software. In the next section, we discuss the phase transitions in the model.

\begin{figure}[t]
	\centering
	\includegraphics[width=.49\textwidth]{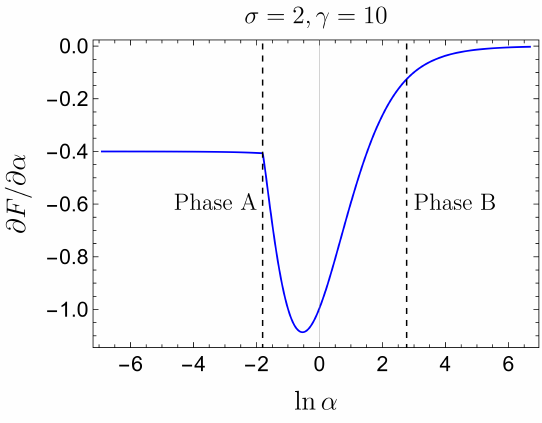}
	\,
	\includegraphics[width=.49\textwidth]{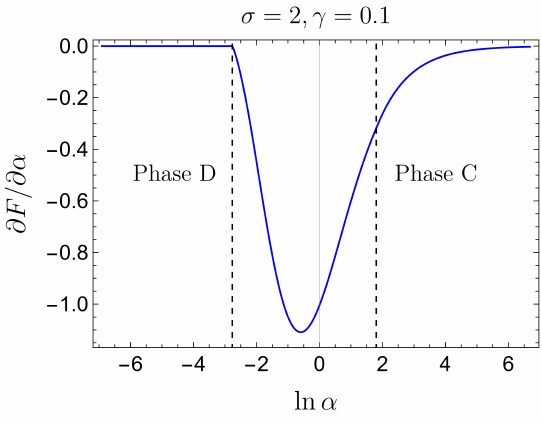} \\ \quad \\ 
	
	\begin{subfigure}[t]{0.49\textwidth}
		\centering
		\begin{tikzpicture}
			\node (main) at (0,0) 
			{\includegraphics[width=\textwidth]{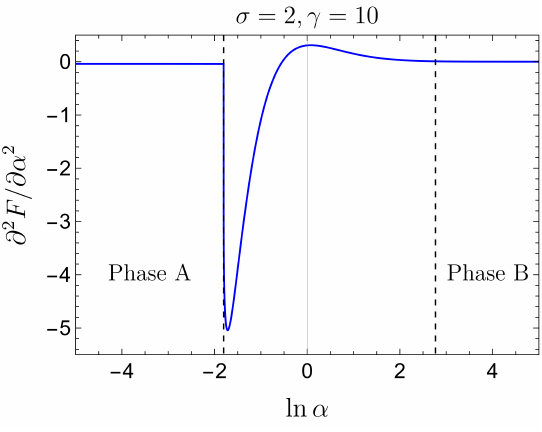}};
			
			\node[anchor=south west] at (0.47,-0.35)
			{\includegraphics[width=0.41\textwidth]{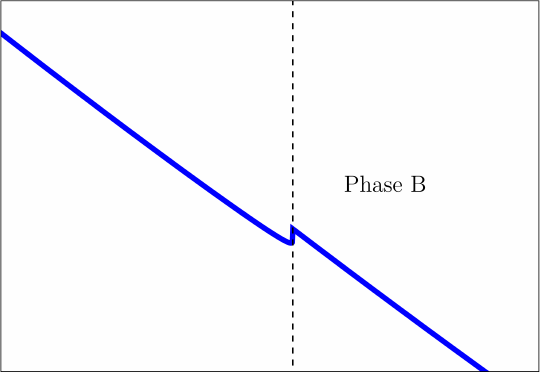}};
		\end{tikzpicture}
		
	\end{subfigure}
	\hfill
	\begin{subfigure}[t]{0.49\textwidth}
		\centering
		\begin{tikzpicture}
			\node (main) at (0,0) 
			{\includegraphics[width=\textwidth]{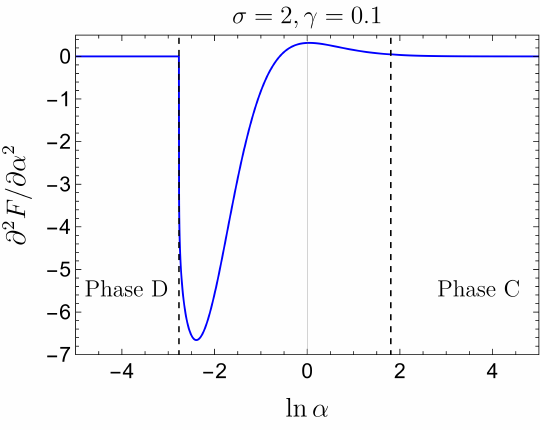}};
			
			\node[anchor=south west] at (-0.14,-0.35)
			{\includegraphics[width=0.41\textwidth]{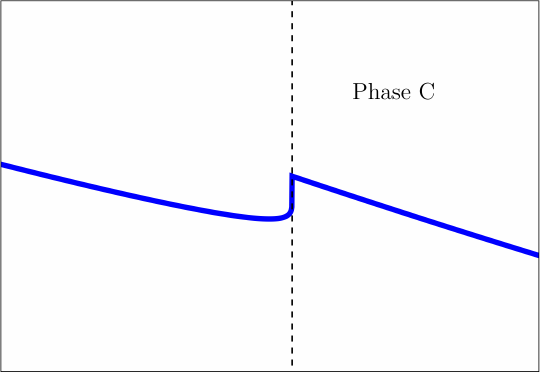}};
		\end{tikzpicture}
		
	\end{subfigure}
	\caption{\small The first two derivatives of the large-$N$ free energy are continuous at the transition point. However, the second derivative of the free energy has a kink at the transition points, which signals a third order transition. The inset in the bottom panels zooms in on the kink.}
	\label{fig:freeEnergyThroughTransition}
\end{figure}

\section{Summary of phase transitions} \label{sec:PhaseTransitions}

This model, similar to the Gross-Witten-Wadia model, exhibits third order phase transitions in the large-$N$ limit. See Fig.~\ref{fig: Phase diagram} for the full large-$N$ phase diagram. The present model addtionally exhibits finite-$N$ phase transitions whenever the inverse of the coupling is a non-integer, due to the non-analyticity of the potential at the branch cuts. The order of these finite-$N$ transitions depends on the coupling $g$ and is given by $1 + \mathrm{ceiling}(1/g)$. This fact ensures that the transitions disappear in the double scaling large-$N$ limit, and they are replaced by an ungapped-to-gapped transition. Interestingly, there are no direct transitions between the different ungapped phases. We also note that as the coupling $\sigma$ increases, the gapped phase expands outwards from the center, while the ungapped phases shrink. At $\sigma=0$ a single ungapped phase exists, while at $\sigma = \infty$ only the gapped phase remains. All five phases exist at every finite value of $\sigma$. Fig. \ref{fig:gappedWilson} shows the Wilson loop expectation values and the Lagrange multiplier as the system transitions from one ungapped phase to another through the intermediate gapped phase. The Wilson loop $W_1$ in particular exhibits a behaviour similar to the famous Silver Blaze phenomenon of QCD, where a thermodynamic quantity is independent of a parameter (the chemical potential to be precise) on one side of the transition, and dependent on the other, despite the fact that the parameter appears explicitly in the action.

In the large-$N$ limit, the second derivative of the Lagrange multiplier $\mathscr{N}$ and the third derivative of the free energy are discontinuous at the transition, indicating that the phase transition is third order. See Fig.~\ref{fig:freeEnergyThroughTransition} for plots of the derivatives of $F$. Although derivatives of $F$ with respect to only $\alpha$ are shown, the discontinuity in the third derivative $F$ exists for all parameters, including combinations such as $\frac{\partial^3 F}{\partial \alpha \, \partial^2 \gamma}$.

\begin{figure}[t]
	\centering
	\includegraphics[width=.49\textwidth]{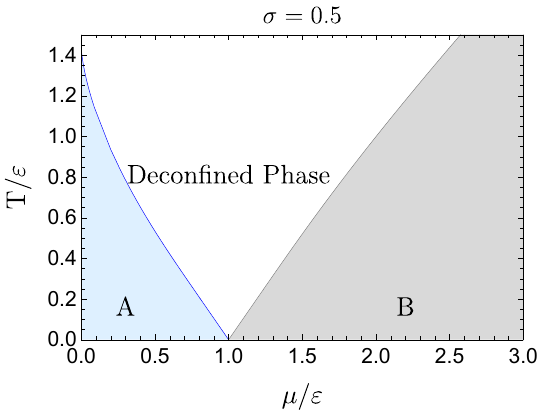}
	\,
	\includegraphics[width=.49\textwidth]{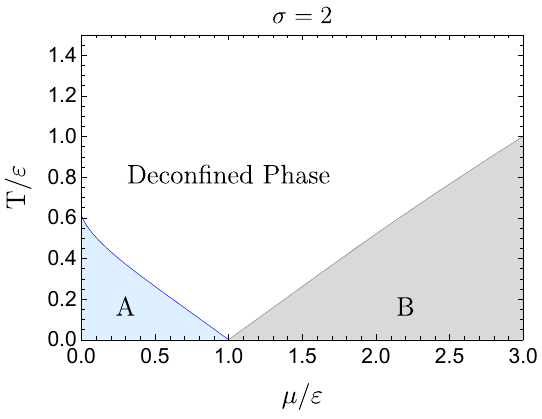}
	\caption{\small $T$-$\mu$ phase diagram of the large-$N$ model when related to weakly-coupled QCD. There are two confined phases (colored) and one deconfined phase.}
	\label{fig:muTPhaseDiagram}
\end{figure}

Finally, as mentioned in the introduction, our model converges to the low temperature two-level QCD model of Ref. \cite{hands_qcd_2010} if we relate the parameters $\alpha, \gamma$ to the temperature $T$ and the chemical potential $\mu$ as follows
\begin{equation} \label{eq:alphaGammaMapping}
    \alpha = e^{(\mu - \varepsilon)/T}, \qquad \gamma = e^{(\mu + \varepsilon)/T} 
\end{equation}
where $\varepsilon$ is the energy difference between the two levels. In this case, $\sigma$ is proportional to the number of quark flavors, $\mathscr{N}$ becomes the fermion number. In this interpretation, the finite-$N$ transitions occur at $\mu_c = \varepsilon$. The large-$N$ ungapped and gapped phases have the interpretation of confined and deconfined phases. In the exact zero-temperature limit $\gamma$ approaches infinity, and one of the log terms in the potential \eqref{def:potential} can be dropped, which is exactly what the authors of Ref. \cite{hands_qcd_2010} have done. Keeping both terms produces the phase diagram shown in Fig.~\ref{fig:muTPhaseDiagram}, which agrees with that in Ref. \cite{hands_qcd_2010} at small temperatures. Only two confined phases exist in this diagram, as the other two correspond to negative temperature. We also note that as $\sigma$ increases, the deconfined phase expands, while the confined phase shrinks to smaller values of $T$. The three phases exist at every finite value of $\sigma$. Lastly, it appears in the plots that a direct confined to confined transition is allowed at $T=0$. This apparent contradiction with the previous discussion is resolved by noting that the potential \eqref{def:potential} does not really cover the $T=0$ case. Specifically, the mapping to the parameters $\alpha$ and $\gamma$ given in \eqref{eq:alphaGammaMapping} is pathological at $T=0$.

\section{Conclusion} \label{sec:conclusion}

In this work, we have examined an SU($N$) unitary matrix model with a complex action containing Fisher-Hartwig singularities. The model exhibits phase transitions of coupling-dependent order at finite-$N$. These transitions disappear in the exact large-$N$ limit, and replaced by third order Gross-Witten-Wadia transitions between four ungapped phases and a single gapped phase. Transitions between the ungapped phases proceed through the intermediate gapped phase, never directly. The saddle points in the large-$N$ limit were found to lie off the unit circle in the complex plane for both ungapped and gapped phases, reflecting the inherent complex-valued nature of the model's action.

Exact and explicit expressions were obtained for all observables in the ungapped phases, whereas in the gapped phase the transcendental nature of the equations forced us to use numerical methods to study the observables. We have also connected our results to an existing model of low-temperature weak-coupling QCD to verify our results, and extended the existing results.

These results provide deeper insight into the phase structure of complex extensions of unitary matrix models with Fisher-Hartwig singularities and may offer useful analogies for understanding complex actions in QCD-like theories. Further extensions of this work could include complete expressions for the finite-$N$ observables and their detailed anaylsis, or exploring the behavior under further complex deformations of the potential.

\bibliographystyle{plain}
\bibliography{references}

\end{document}